\documentclass[11pt,preprintnumbers,pra,aps,
nofootinbib,tightenlines,floatfix, twocolumn,superscriptaddress,longbibliography]{revtex4-2}

\def\Nslash{{ {\cal N}\hskip-0.55em /}}

\usepackage[colorlinks=true]{hyperref}
\hypersetup{
  linkcolor  = TealBlue!90!Black,
  citecolor  = YellowGreen!90!Black,
  urlcolor   = Purple!90!Black,
  colorlinks = true,
}
\usepackage[dvipsnames]{xcolor}
\usepackage[all]{hypcap}
\usepackage{amsmath,amssymb}
\usepackage{graphicx}
\usepackage{physics}
\usepackage{balance}
\usepackage[utf8]{inputenc}
\usepackage{csquotes}
\usepackage{empheq}
\usepackage{longtable}
\usepackage{multirow}
\usepackage[section]{placeins}
\usepackage{comment}
\usepackage[customcolors]{hf-tikz}
\usepackage[top=1in,bottom=1in,left=1.8cm,right=1.6cm]{geometry}

\DeclareMathOperator{\diag}{\text{diag}}

\allowdisplaybreaks
\makeatletter
\renewcommand\onecolumngrid{
\do@columngrid{one}{\@ne}
\def\set@footnotewidth{\onecolumngrid}
\def\footnoterule{\kern-6pt\hrule width 1.5in\kern6pt}%
}
\makeatother

\newcommand\scalemath[2]{\scalebox{#1}{\mbox{\ensuremath{\displaystyle #2}}}}

\newcommand*\pFqskip{8mu}
\catcode`,\active
\newcommand*\pFq{\begingroup
        \catcode`\,\active
        \def ,{\mskip\pFqskip\relax}%
        \dopFq
}
\catcode`\,12
\def\dopFq#1#2#3#4#5{%
        {}_{#1}F_{#2}\biggl[\genfrac..{0pt}{}{#3}{#4};#5\biggr]%
        \endgroup
}

\newcount\hour \newcount\hourminute \newcount\minute
\hour=\time \divide \hour by 60
\hourminute=\hour \multiply \hourminute by 60
\minute=\time \advance \minute by -\hourminute

\begin{document}

\title{Detecting spacelike vacuum entanglement at all distances and promoting negativity to a necessary and sufficient entanglement measure in many-body regimes}

\author{Boyu Gao}
\email{boyu.gao@duke.edu}
\affiliation{{Duke Quantum Center and Department of Physics, Duke University, Durham, NC 27708, USA}}
\author{Natalie Klco}
\email{natalie.klco@duke.edu}
\affiliation{{Duke Quantum Center and Department of Physics, Duke University, Durham, NC 27708, USA}}

\begin{abstract}

Though known to be present, the accessibility of spacelike vacuum entanglement capable of being a fundamental resource for quantum information processing has remained in question at distances beyond the scale of vacuum fluctuations in massive fields. For a broad subclass of physical many-body mixed Gaussian states, including the free scalar field vacuum, the logarithmic negativity is here shown to be a necessary and sufficient measure of entanglement and to be entirely accessible by pairs of single-mode detectors in the continuum. By deriving exact and optimal detection profiles, entanglement resources in the massive field are demonstrated to be available at all distances.  

\end{abstract}
\date{\today}
\maketitle

{
\footnotesize
\tableofcontents
}

\section{Introduction}
\label{sec:sec1}

Quantum field vacuums naturally distribute quantum entanglement~\cite{EPRoriginal,Bellineq} at spacelike separations~\cite{ReehSchlieder,SUMMERS1985257,summers1987maximal,summers1987bell1,summers1987bell2,Witten}, i.e., quantum information not only propagates within but is an essential element of quantum fields themselves. With the importance of distributed entanglement arising throughout scientific frontiers---from black hole physics~\cite{blackhole1986,Srednicki:1993im} to the development of quantum information processing technologies, e.g., for simulation and sensing~\cite{Degen,Banuls,coldatom,Marciniak,NKreview}---it is valuable to understand the information-theoretic relationship between spacetime and entanglement. One insightful approach providing operational perspective has been utilizing pairs of small quantum systems as external detectors that extractively probe spatial quantum correlations via interaction with local field regions~\cite{Unruh1,Valentini,vaneg2,Reznik1,Reznik2,Hu:2012jr,vaneg6,Martin-Martinez:2015qwa,Hackl:2019hfl}. Serving as a mixed quantum information resource, where the logarithmic negativity~\cite{peresoriginalN,HORODECKIoriginalN,Simonreflection,computablemeasure, PlenioLogarithmic} continues to grow in importance for quantifying field entanglement~\cite{scalarvacuumoriginal1,onebeforeZpaper,scalar1dextra,vaneg5,Zpaper,vaneg6,calabrese2014finite,Wen:2015qwa,Kudler-Flam:2018qjo,Kusuki:2019zsp,NKnegativitysphere,NKentsphere,missep1PRD2023,agullo2024multimodenaturespacetimeentanglement,missep1PRR2024}, these regions reside in the challenging regime of many-body entangled states~\cite{Fourqubits,It2,Huang_2014} extending to infinite-body for continuous fields. Despite the inseparability of such continuum regions,  various critical distances around the scale of massive vacuum fluctuations have been reported, through both algebraic approaches~\cite{Reznik1,missep1PRR2024} and lattice collective operators~\cite{onebeforeZpaper,Zpaper,missep1PRD2023}, beyond which it is asserted that no operational entanglement resources are available. Leveraging the clarity of lattice regularization and the quantum information structure provided by local symmetry transformation to the partially transposed (PT) eigenspace, we derive collective detection modes for accessing field entanglement at all spatial separations in the free massive continuum field.

The massive field exhibits novel functional forms of spatially distributed entanglement~\cite{scalar1dextra} and quantum coherent lattice volumes. However, the exponential relationship between the UV spatial resolution and the smallest spacelike entanglement supported by the lattice, i.e., $\mathcal{N} \sim e^{-\Lambda'd}$ we observe to persist from the massless regime~\cite{NKentsphere}, allows rigorous determination that the physical space of entangled vacuum extends to infinity in the continuum limit.

Expressed in the advantageous language of Gaussian quantum information~\cite{Duanlocaltrans,Simonreflection,giedke2000inseparable,giedkedistill,It1,BraunsteinGaussiareview,It3,horodecki2009quantum,WeedbrookGaussiareview,It3,serafini2017quantum,Hackl:2020ken}, spacelike entanglement within the latticized free field vacuum in the thermodynamic limit of infinite volume is entirely characterized by the Covariance Matrix (CM) of the detection regions. By leveraging translational invariance to establish semidefinite properties of the region-region CM, we determine that logarithmic negativity~\cite{peresoriginalN,HORODECKIoriginalN,Simonreflection,computablemeasure,PlenioLogarithmic} is necessary and sufficient for identifying this field entanglement and that it is amenable to reorganization into an exponential hierarchy of $(1_A \times 1_B)$ entangled pairs via local operations. With these properties applying beyond the free scalar field vacuum, this work extends the direct significance of the negativity entanglement measure from low dimensional or highly symmetric systems---e.g., two-mode, $(1_A \times n_B)$-mode, pure, isotropic, or bisymmetric states~\cite{peresoriginalN,HORODECKIoriginalN,Simonreflection,PhysRevA.67.052311,GiedkeEOF,WolfGEOF,serafini2017quantum}---to a structural class of physical many-body Gaussian quantum states.

In addition to the immediate fundamental physics of identifying the availability of vacuum entanglement and informing the design of detectors capable of connecting to the field as a quantum resource for subsequent quantum information processing, it is anticipated that this understanding will contribute to the entanglement-guided design of quantum simulation algorithms~\cite{feynman2018simulating,NKdesign,simulationdesignalter,Robin:2023pgi,PhysRevLett.132.100402,NKphononatural,friesecke2024globalfermionicmodeoptimization}, e.g., for applications in nuclear and particle physics.

This article is organized as follows. Section~\ref{sec:sec2} presents an overview of Gaussian states and their application to the scalar field vacuum, including the addition of several semidefinite properties of field region-region CMs derived analytically in Appendix~\ref{app:fieldintro}. Section~\ref{sec:newsec3} demonstrates that, for the general class of states exhibiting these properties, the PT criterion becomes both necessary and sufficient for identifying entanglement, and the PT space is sufficient to identify local symplectic transformations that reorganize negativity into entangled pairs. The detailed proofs supporting this section are provided in Appendix~\ref{app:struc}. Section~\ref{sec:sec3} builds on this understanding to construct optimal collective detector profiles capable of transferring spacelike field entanglement into a pair of single-mode detectors, as illustrated in Fig.~\ref{fig:localsprowvec}.  The proof of optimality is provided in Appendix~\ref{app:opt}. Finally, Sec.~\ref{sec:sec4} synthesizes the above advances to show that spacelike entanglement remains accessible at all separations in the continuum limit, supported by the scaling behavior shown in Fig.~\ref{fig:heatmap} and Fig.~\ref{fig:plots} of Appendix~\ref{app:artfact}.

\section{Vacuum entanglement}
\label{sec:sec2}

\begin{figure*}
 \begin{minipage}{0.95\textwidth}
 \includegraphics[width = \textwidth]{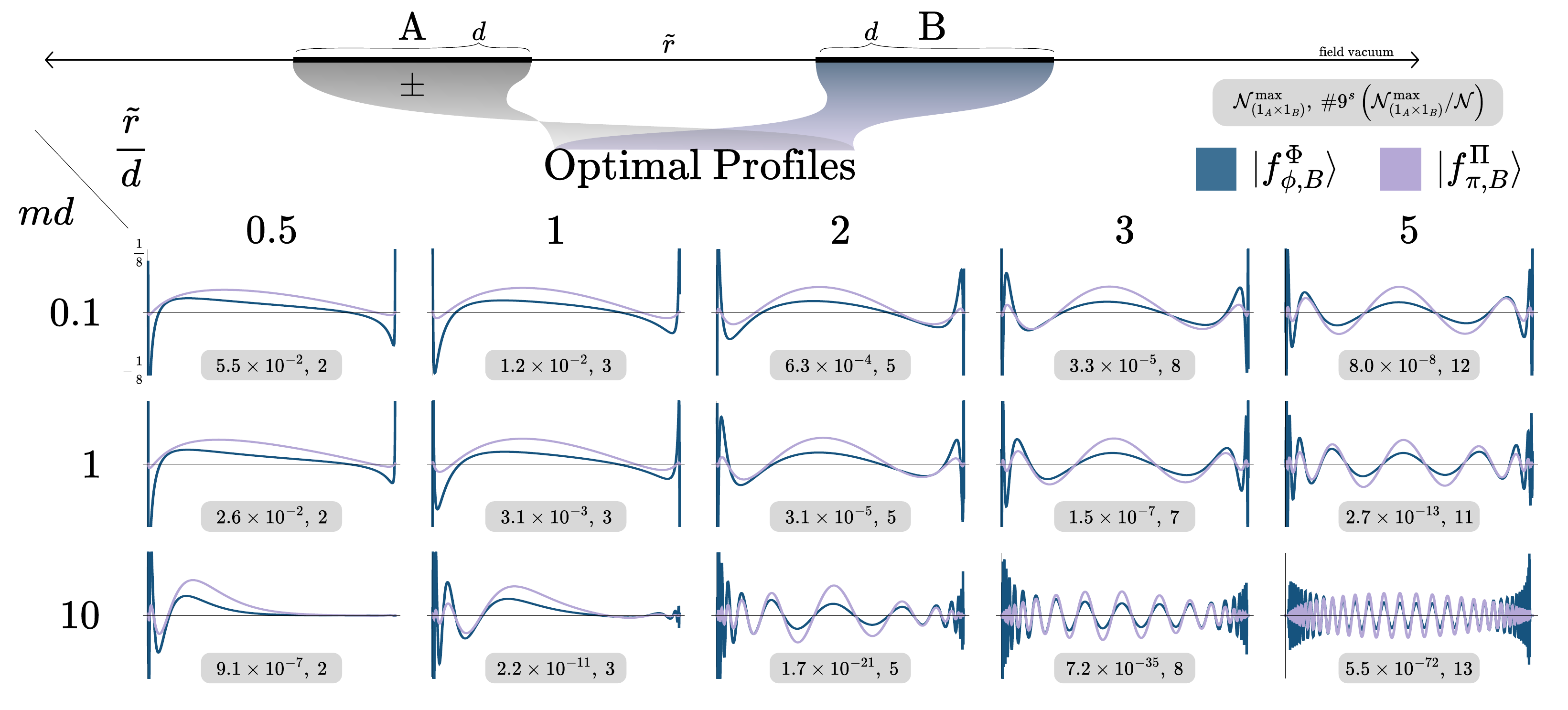}
  \end{minipage}\\
  \caption{Optimal mode profiles for connecting to spacelike entanglement via local detectors in the massive scalar field vacuum at various configurations, $md$ and $m\tilde{r}$, measuring the regions' size and separation in units of the Compton wavelength scale. Profiles in the A-space relate to those depicted for the B-space via reflection and (optional) negation. For each configuration, values are provided for the maximum logarithmic negativity $\mathcal{N}^{\text{max}}_{(1_A \times 1_B)}$ extractable by the indicated pair of detection modes and the number of nines in its ratio with the total available entanglement between the field regions. The pixelation of $d=1000$ oscillators per region provides a close approximation to the continuum. These equal-norm $\phi\pi$-uncorrelated interactions each represent a family of optimal profiles defined by a flexibility of local operations on the collective mode.}
  \label{fig:localsprowvec}
\end{figure*}

The free scalar field vacuum is a Gaussian state, which can be fully described by first moments, $\bar{\boldsymbol{r}}$, of phase space operators and their CM, $\sigma = \mathrm{Tr} \left[ \rho \left\{ (\hat{\boldsymbol{r}} - \bar{\boldsymbol{r}}),(\hat{\boldsymbol{r}} - \bar{\boldsymbol{r}})^{T} \right\}  \right]$, where $\rho$ is the density matrix. Capturing the entanglement information, the lattice scalar field vacuum CM, $\sigma = \sigma_\phi \oplus \sigma_\pi$, has vanishing field-momentum matrix elements (see Appendix~\ref{app:fieldintro} for detailed discussion). For analytically approaching many aspects of the entanglement structure between disjoint vacuum regions, it is convenient to consider the vector of phase-space operators, numbered from left-to-right in one spatial dimension, organized in the following vector order,
\begin{equation}
\scalemath{0.99}{\hat{\boldsymbol{r}} = \left( \phi_{A_{d}}...\phi_{A_{1}},\phi_{B_{1}}...\phi_{B_{d}},\pi_{A_{d}}...\pi_{A_{1}},\pi_{B_{1}}...\pi_{B_{d}}\right)^{T}}    ,
 \label{eq:orderofcan}
\end{equation}
where $d$ is the number of lattice sites per region and $\tilde{r}$ is the number of lattice sites between $\phi_{A_d}$ and $\phi_{B_1}$ as depicted in Fig.~\ref{fig:localsprowvec}. In this ordering, the symplectic matrix of canonical commutation relations (CCRs) takes the form, $\Omega_{2d} = -i\left[ \hat{\mathbf{r}}, \hat{\mathbf{r}}^T \right] = \begin{pmatrix} 0 & \mathbb{I}_{2d} \\ -\mathbb{I}_{2d}&0\end{pmatrix}$. Beyond the $\sigma_{\phi,A} = \sigma_{\phi,B}$ and $\sigma_{\pi,A} = \sigma_{\pi,B}$ symmetry, this ordering also builds-in several advantageous properties, including semi-definiteness of field and momentum space CM region-region (off-diagonal) blocks,
\begin{equation}
\sigma_{\phi,AB} \geq 0, \quad \sigma_{\pi,AB} \leq 0  \ \ \ ,
\label{eq:offblockmatcondition}
\end{equation}
which we show to be analytically satisfied in Appendix~\ref{app:fieldintrooff}. The CM can always be diagonalized to Williamson normal form~\cite{williamsonnormalform} via symplectic transformation~\footnote{Constructive procedures for calculating an $S$ that performs this symplectic diagonalization are discussed, for example, in Appendix D of Ref.~\cite{NKvolumemeasure} or Appendix B2 of Ref.~\cite{gao2024partialtransposeguided}.}, $ S \sigma S^{T} = D$, where D is a diagonal matrix of doubly-degenerate real symplectic eigenvalues. For a physical CM, $D\geq \mathbb{I}$, where $\mathbb{I}$ denotes the identity matrix. 

The Peres-Horodecki criterion~\cite{peresoriginalN,HORODECKIoriginalN,Simonreflection} states that the presence of negative, i.e., unphysical, eigenvalues in a partially transposed (PT) density matrix is a sufficient condition for quantum systems of any number of modes to be entangled. For Gaussian states, this entanglement criterion is associated with the physicality of the PT CM, denoted as $\Tilde{\sigma} = \Lambda \sigma \Lambda$, where $\Lambda_{d}=(\mathbb{I}_{d} \oplus \mathbb{I}_{d} \oplus \mathbb{I}_{d} \oplus -\mathbb{I}_{d})$ is the momentum reversal operator within the B half-space~\cite{Simonreflection}. Logarithmic negativity~\cite{computablemeasure} is a computable measure of entanglement built upon the Peres-Horodecki criterion. For Gaussian states, it reads,
\begin{equation}
    \mathcal{N} = -\sum_{j=1}^{n_{-}}\log_{2}\Tilde{\nu}_{j} \ \ \ ,
    \label{eq:negg}
\end{equation}
where $n_- \leq d$ denotes the number of PT symplectic eigenvalues, $\tilde{\nu}_j$, that are smaller than one, e.g., obtained from the normal mode decomposition $\tilde{S} \tilde{\sigma}\tilde{S}^T = \tilde{D}$ of the PT CM.

\section{Necessary and sufficient properties}
\label{sec:newsec3}

In the normal form of two-mode Gaussian states~\cite{Simonreflection,Duanlocaltrans}, entangled CMs must have off-diagonal matrix elements of opposite sign, e.g., positive $\langle \phi_1 \phi_2\rangle$ and negative $\langle\pi_1\pi_2\rangle$. Equation~\eqref{eq:offblockmatcondition} generalizes this property to many-body contexts in the form of positive- and negative-semidefinite $\sigma_\phi$ and $\sigma_\pi$ region-region blocks, respectively. For the class of states satisfying these generalized criteria, including and beyond the scalar field vacuum (see Appendix~\ref{app:fieldintrooff}), we derive two significant many-body entanglement properties below. 

For $AB$-symmetric $\phi\pi$-uncorrelated CMs satisfying Eq.~\eqref{eq:offblockmatcondition}, consider a symplectic transformation, $\left(S_0 \oplus S_0\right)$, of the PT CM,
\begin{equation}
    (S_0 \oplus S_0)\tilde{\sigma}(S_0 \oplus S_0)^T \stackrel{\text{reorder}}{=}  \tilde{\sigma}_{\pm} \oplus \tilde{\sigma}_{\mp} \ \ \ ,
    \label{eq:s0andsigmapmmp}
\end{equation}
where $S_{0} \equiv \frac{1}{\sqrt{2}}\begin{pmatrix} \mathbb{I}_{d} & \mathbb{I}_{d} \\ -\mathbb{I}_{d} & \mathbb{I}_{d} \end{pmatrix}$,
\begin{subequations}
\begin{align}
\Tilde{\sigma}_{\mp} &\equiv \left( \sigma_{\phi,A} - \sigma_{\phi,AB} \right) \oplus \left(\sigma_{\pi,A} + \sigma_{\pi,AB} \right)    \\ \Tilde{\sigma}_{\pm} &\equiv \left( \sigma_{\phi,A} + \sigma_{\phi,AB} \right) \oplus \left( \sigma_{\pi,A}  -\sigma_{\pi,AB} \right)  \ \ \ ,
\end{align}
\label{eq:newsub}%
\end{subequations}
and reorder refers to rearranging the canonical operators so that the first and second halves of the $\phi$ and $\pi$ spaces are grouped separately. For such CMs, the Peres-Horodecki criterion~\cite{peresoriginalN,HORODECKIoriginalN,Simonreflection} for the PT CM can be simplified to,
\begin{equation}
\sigma_{\phi,A} - \sigma_{\phi,AB} - \left(\sigma_{\pi}^{-1}\right)_{A} - \left(\sigma_{\pi}^{-1}\right)_{AB} \geq 0 \ \ \ ,
\label{eq:PHcriterionsimp}
\end{equation}
which is equivalent to the physicality condition of $\tilde{\sigma}_{\mp}$. When the negativity vanishes (PPT), Eq.~\eqref{eq:PHcriterionsimp} thus enables the identification of a separable decomposition $\sigma^{sep} \leq \sigma$, where $\sigma^{sep}$ is constructed to be $\tilde{\sigma}_{\mp}$ in both the $A$ and $B$ spaces (see Appendix~\ref{app:struc1} for details). As such, the Peres–Horodecki criterion is both necessary and sufficient for identifying entanglement in this case, e.g., for spacelike regions of the free scalar field vacuum as well as the Gaussian approximation of local axial motional modes symmetrically located in a trapped-ion chain~\cite{NKphononatural}. This result provides analytic support for previous observations of the entanglement sphere coinciding with the negativity sphere~\cite{NKentsphere} via numerical techniques~\cite{Giedkesepflow}, and extends the necessary and sufficient property of the Peres–Horodecki condition to a broad class of many-mode symmetric Gaussian mixed states.

With the negativity contributing subspace ($\mathcal{V_{\mathcal{N}}}$) contained within that of $\tilde{\sigma}_{\mp}$, $AB$-symmetric $\phi\pi$-uncorrelated CMs satisfying Eq.~\eqref{eq:offblockmatcondition} allow local transformations that consolidate negativity (see Appendix~\ref{app:stru2} for details). Such transformations consist of local symplectic operations that reorganize multimode negativity into a tensor-product series of $(1_{A} \times 1_{B})$ pairs~\cite{NKcorehalo,gao2024partialtransposeguided}. This assured consolidation allows all available entanglement to be transferred to pairs of single-mode detectors for both the lattice and continuum field. The isolation of Eq.~\eqref{eq:s0andsigmapmmp} also reduces by a factor of four the matrix dimension required to calculate the negativity and symplectic transformation that diagonalizes the PT CM to normal form (see Appendix~\ref{app:stru3} for details). This provides numerical advantage in the calculation of the available entanglement resource and the relevant local collective operators capable of detecting it.

\section{Collective operators}
\label{sec:sec3}

To design two single-mode detectors of field entanglement between regions of finite size in the continuum, the collective operators of the dominant $(1_A \times 1_B)$ entangled pairs must be identified. Such collective operators are defined as weighted averages over phase-space operators local to each region,
\begin{subequations}
            \begin{align}
             \Phi &= \sum_{j=1}^{d} f_{\phi,j}^{\Phi} \phi_{j} + \sum_{j=1}^{d} f_{\pi,j}^{\Phi} \pi_{j} \\ \Pi &= \sum_{j=1}^{d} f_{\phi,j}^{\Pi} \phi_{j} + \sum_{j=1}^{d} f_{\pi,j}^{\Pi} \pi_{j}
    \ \ \ ,
\end{align}
\label{eq:generalizeZpapaerprofiles}%
\end{subequations}
with normalization $\sum_{j=1}^{d} ( f_{\phi,j}^{\Phi} f_{\pi,j}^{\Pi} - f_{\pi,j}^{\Phi} f_{\phi,j}^{\Pi})  = 1$.  An analogous expression applies to B-space canonical operators. The normalization condition ensures that the collective two-mode operator space, $\hat{\mathbf{r}}_f \equiv  \left( \Phi_A , \Phi_B , \Pi_A , \Pi_B \right)^{T}$, continues to satisfy CCRs $ \left[ \hat{\mathbf{r}}_f, \hat{\mathbf{r}}^T_f\right] = i \Omega_2$. Optimal identification of these collective degrees of freedom can reduce the many-body disjoint vacuum regions to pairs of bosonic modes with an exponential hierarchy in their entanglement~\cite{NKcorehalo,gao2024partialtransposeguided}.

To calculate collective operators that detect the maximum possible two-mode logarithmic negativity in the vacuum, Appendix~\ref{app:profiles} begins by relating profiles of collective operators to row vectors of arbitrary local symplectic transformations, 
\begin{equation}
    S\sigma S^{T}= \mathrm{Tr} \left[ \rho \left\{ S\hat{\boldsymbol{r}},\hat{\boldsymbol{r}}^{T} S^{T} \right\}   \right] \ \ \ .
\label{eq:eq4}
\end{equation}
The connection of Eq.~\eqref{eq:eq4} leads optimal profiles to be calculated from row vectors of the local consolidation transformation~\cite{NKcorehalo,gao2024partialtransposeguided} discussed in Sec.~III. In particular, the optimal profile corresponds to the dominant negativity contribution. This is distinct from the procedures in, for example, Ref.~\cite{Zpaper} where field- and momentum-space profiles are assumed to be equal, and a numerical optimization program is designed~\footnote{Note that the equal-profile framework~\cite{onebeforeZpaper,Zpaper,missep1PRD2023,agullo2024multimodenaturespacetimeentanglement}, constraining Eq.~\eqref{eq:generalizeZpapaerprofiles}, is found to be responsible for prohibiting the identification of entanglement and associated collective operators beyond the Compton wavelength scale.}.

For disjoint regions of the free scalar vacuum, these detection profiles can exhibit no $\phi\pi$-mixing (i.e., $f_{\phi,j}^\Pi = f_{\pi,j}^\Phi = 0$) and the $(4d \times 4)$ relevant component of the symplectic operator becomes ${S_f^T \equiv \diag \left(|f^\Phi_{\phi,A}\rangle, |f^\Phi_{\phi,B}\rangle, |f^\Pi_{\pi,A}\rangle, |f^\Pi_{\pi,B}\rangle\right)}$, with several optimal profiles illustrated in Fig.~\ref{fig:localsprowvec}. Finally, because regions of the continuum field have $n_- >1$ negativity contributions and Appendix~\ref{app:opt} establishes an upper bound on the $(1_A \times 1_B)$ consolidated entanglement governed by the minimum PT symplectic eigenvalue of the multi-mode system, the entanglement of the field vacuum is found to have a fundamental incompressibility, i.e., these profiles that saturate the upper bound access an exponentially dominant component, but never the entirety, of the entanglement.

Governed by these collective operator profiles, extracting vacuum entanglement may be achieved via a beamsplitter interaction, $H_{BS}=\Pi \phi_{D} -\Phi \pi_{D}$ that generates the symplectic transformation $S_{BS}=e^{\Omega H_{BS} \theta}$~\cite{serafini2017quantum}. With the choice $\theta=\pi/2$, this interaction swaps the single-mode detector, $\phi_{D}$ and $\pi_{D}$, with the collective field degree of freedom, $\Phi$ and $\Pi$. By performing such a process in each local region, entanglement distributed over many field modes may be transferred to a pair of detectors, leading towards the experimental detection of spacelike vacuum entanglement.

\section{Accessing entanglement at all distances}
\label{sec:sec4}

\begin{figure}
  \includegraphics[width =\columnwidth]{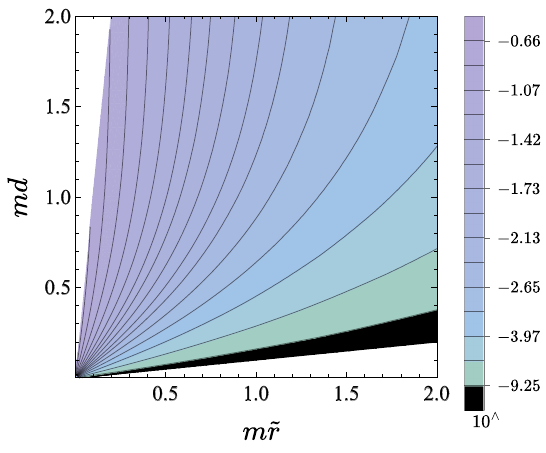}
  \caption{Logarithmic negativity $\mathcal{N}$ between disjoint regions of the massive free scalar field vacuum for select configurations governed by $md$ and $m\tilde{r}$. Contour lines correspond to values of $\mathcal{N}$ in the massless regime with $\tilde{r}/d$ and $d/\tilde{r}$ ratios of {1, 7/6, 4/3, 8/5, 2, 8/3, 4, 7}.   Values are calculated with pixelations $d \gtrsim 100$ for ratios with minimal lattice configurations $\tilde{r}_{\text{min}}, d_{\text{min}} \leq 10$, i.e., the white regions indicate absent calculations not vanishing $\mathcal{N}$.}
  \label{fig:heatmap}
\end{figure}

With entanglement between vacuum regions vanishing at finite separations in the presence of a UV truncation~\cite{NKentsphere}, the range of continuum entanglement is determined from the latticized field by observing the expansion of such entanglement spheres toward the continuum limit. For massless fields, this expansion of the coherent volume is approximately linear in the lattice resolution, whereas this growth (and convergence to the continuum entanglement) slows with increasing mass. Though the direct calculations of Appendix~\ref{app:entsphere} indicate that the massive entanglement sphere indeed expands beyond the scale of the Compton wavelength, the reduced rate of its expansion makes it numerically challenging to assure that growth continues indefinitely as the continuum is approached. To rule out saturating behavior, first the minimum negativity supported by the lattice is observed to scale exponentially with the UV truncation, $\mathcal{N}_{\Nslash} \sim e^{-\Lambda' d}$, a relationship persisting from the massless regime~\cite{NKentsphere}. Combined with the quadratic exponential decay $\mathcal{N} \sim e^{-\left(m\tilde{r}\right)^2}$~\cite{scalar1dextra} at large separations, extending the relationship of long-distance UV sensitivity into the massive regime (also visible in the increasing profile oscillation frequency with separation $\tilde{r}/d$ at the right of Fig.~\ref{fig:localsprowvec}) allows the deduction of a square root functional form for the expansion of the massive field entanglement sphere with increasing UV resolution. As such, the lattice regularization offers clear indication that finite spacelike entanglement is systematically supported at all separations in the continuum field.  

The non-vanishing value of continuum logarithmic negativity available between vacuum regions is shown in Fig.~\ref{fig:heatmap} throughout the two-dimensional configuration space of $md$ and $m\tilde{r}$, with the latter separations surrounding the Compton wavelength scale. The contour lines begin radially in the massless regime where the negativity depends only on $\tilde{r}/d$~\cite{scalar1dextra,NKnegativitysphere,NKentsphere}, curve up as the mass causes the entanglement to decay more rapidly, and become parallel for $md >>1$ where the exponential decay of the massless regime $\mathcal{N}\sim e^{-\tilde{r}/d}$ translates to a regime of $\mathcal{N}\sim e^{-mr}$ before the quadratic exponential decay for $m \tilde{r} \gtrsim md$~\cite{scalar1dextra}. Because the $\mathcal{N}$-SOL classification of the lattice and continuum scalar field (see Appendix~\ref{app:stru2}) allows the logarithmic negativity to reflect the amount of entanglement transferable to pairs of single-mode detectors~\cite{gao2024partialtransposeguided}, these calculations visually confirm that no truncation or discontinuity is present in the entanglement resource for separations at the scale of the Compton wavelength.

\section{Discussion}
\label{sec:sec5}

By designing a local Hamiltonian interaction with field collective operators, this manuscript demonstrates that entanglement between disjoint regions of continuum scalar vacuum is always available to be extracted with a pair of single-mode detectors as a viable quantum correlation resource. The profiles that maximize the detected entanglement are shown to be directly calculable from the substructure of the PT symplectic eigensystem. With long distance entanglement requiring systematically higher frequency fluctuations in the optimal profile, the distance at which entanglement may be accessed from the continuum scalar vacuum is thus governed only by the spatial resolution of the detector's connection to the field. In general, local probes of spacelike entanglement are anticipated to share this feature.

While deeper understanding of the functional forms of optimal profiles remains to be uncovered, present calculations have identified their key features, including independence of the field and conjugate-momentum space couplings for accessing entanglement at distances beyond the scale of the Compton wavelength. Though visually suggested by similarities in Fig.~\ref{fig:localsprowvec}, utilizing profiles from alternate configurations with negativity values that match to high precision commonly results in separability between the two collective modes. This is one example revealing considerable sensitivity in the success of entanglement detection on the form of the interaction profile. As such, entanglement renormalization perspectives may be valuable in designing approximations to derived profiles compatible with experimental tolerances.

The present time-independent protocol for swapping quantum states between a single-mode detector and a collective mode of the local field region may also guide the design of a time-dependent interaction if working with naturally dynamical fields rather than in the context of information processing.

Through detailed dialogue with the spacelike entanglement distributed in scalar field vacuum, a broad class of Gaussian quantum states has here been identified for which the negativity is promoted to a necessary and sufficient entanglement measure. Beyond the field, this class includes collections of leading-order local axial motion in a trapped-ion chain, motivating an expectation that the visibility of entanglement via subsystem momentum reversal (partial transposition) plays a more profound role in physical quantum systems than previously appreciated for its computational convenience. The observed promotion further accounts for the significant guidance provided by the PT space in entanglement consolidation~\cite{NKcorehalo,gao2024partialtransposeguided}. While the PT space yields complete clarity in determining optimal collective operators for accessing spacelike field entanglement, it is worth noting that open questions---e.g., the conjectured exponential decay of the Gaussian entanglement of formation between spacelike field regions~\cite{gao2024partialtransposeguided}---remain that have thus far eluded determination via PT perspectives and may require new techniques, e.g., inspired by strategic filtration~\cite{Giedkesepflow,gao2024partialtransposeguided} of the classical noise experienced by local observations.

\begin{acknowledgments}

For interactions during the creation of this manuscript, we thank D. H. Beck, I. Marvian, and participants of the \emph{Informational Foundations of QFT} workshop supported by the Wallenberg Initiative on Networks and Quantum Information (WINQ) at Nordita. We thank also participants of the \emph{Entanglement in Many-Body Systems: From Nuclei to Quantum Computers and Back} workshop at the InQubator for Quantum Simulation (IQuS) hosted by the Institute for Nuclear Theory (INT) for interactions at the culmination of this work. IQuS is supported by U.S. Department of Energy, Office of Science, Office of Nuclear Physics, under Award Number DOE (NP) Award DE-SC0020970 via the program on Quantum Horizons: QIS Research and Innovation for Nuclear Science, and by the Department of Physics, and the College of Arts and Sciences at the University of Washington. BG is supported in part by the Goshaw Family Endowment fellowship. Calculations utilizing up to several hundred digits of precision were performed with the Mathematica 14 arbitrary precision libraries~\cite{mma}. Numerical values presented in figures throughout this manuscript are provided in attached files.

\end{acknowledgments}

\bibliography{biblio}

\onecolumngrid
\appendix

\section{Structures of lattice correlation functions}
\label{app:fieldintro}

In this appendix, lattice correlation functions in the thermodynamic limit are discussed in order to analytically derive positivity properties for blocks of the field- and conjugate momentum-space CM, which will be key to the entanglement features of Appendix~\ref{app:struc}.

With discretized field and conjugate momentum operators, $\phi$ and $\pi$, the latticized one-dimensional free scalar field Hamiltonian may be written with unit lattice spacing as,
\begin{equation}
    H =  \frac{1}{2} \sum_{j}  \pi_{j}^{2}   + \frac{1}{2} \sum_{i,j} \phi_{i} ( m^{2} \mathbb{I} -\nabla^2)_{ij} \phi_{j} \ \ \ ,
    \label{eq:descreteH}
\end{equation}
where $\mathbb{I}$ and $\nabla^2$ are the identity and discrete Laplace operator. For example, a leading-order periodic boundary condition (PBC) finite-difference representation of the Laplacian is $\nabla^2 = \text{circulant}\left[ -2, 1, 0, 0, \ldots, 0, 1\right]$, where the content of the bracket indicates the first row or column of a circulant matrix. CCRs for canonical operators ordered as in Eq.~\eqref{eq:orderofcan} can be described with symplectic matrix $\Omega = \begin{pmatrix}
0 & \mathbb{I} \\
-\mathbb{I} & 0 \\
\end{pmatrix}$ such that $[\hat{\boldsymbol{r}},\hat{\boldsymbol{r}}^{T}]=i\Omega$, which is invariant under arbitrary symplectic transformation $S$ with $S \Omega S^{T} = \Omega$. The potential term of Eq.~\eqref{eq:descreteH} can be written in the form of decoupled harmonic oscillators,
\begin{equation}
O_{H}(m^{2} \mathbb{I} -\nabla^2)O_{H}^{T} = D_{H}  \ \ \ ,
\end{equation}
where $O_{H}$ is the orthogonal matrix of eigenvectors that diagonalizes $m^{2} \mathbb{I} -\nabla^2$ and $D_{H}$ is a diagonal matrix of associated eigenvalues. The full volume (pure) free scalar field vacuum CM therefore reads $\sigma_{vac}  = S_{H}^{-1} S_{H}^{-T} \equiv K^{-1} \oplus K$ with,
\begin{equation}
    S_{H} =\begin{pmatrix}
D_{H}^{1/4}O_{H} & 0 \\
0 & D_{H}^{-1/4}O_{H} \\
\end{pmatrix} \ \ \ ,
\end{equation}
such that $K=\sqrt{m^{2} \mathbb{I} -\nabla^2}$, which is analytic when utilizing PBC circulant matrices for the lattice representation.

In the thermodynamic limit of infinite volume, the one-dimensional lattice field and conjugate momentum correlation functions may be written in an integral form, simplified following, for example, Appendix A of Ref.~\cite{NKentsphere},
\begin{align}
    \left(K^{-1}\right)_{0,n} = 2\bra{0}\phi_0\phi_{n}\ket{0}  &= \frac{1}{2\pi} \int_{-\pi}^{\pi} \frac{e^{ipn} dp}{\sqrt{m^2+4\sin^{2}{\left( \frac{p}{2} \right)}}} \\  &= \frac{1}{\sqrt{\pi}(2+m^2)^{n+1/2}}\frac{\Gamma\left[ \frac{1}{2} +n \right]}{\Gamma\left[ 1 +n \right]}  \pFq{2}{1}{\frac{1+2n}{4}\,,\frac{3+2n}{4}}{1+n}{\frac{4}{(2+m^2)^2}} \ \ \  ,
\label{eq:fieldhyp} 
\end{align}
\begin{align}
    K_{0,n} =2\bra{0}\pi_0\pi_{n}\ket{0} &= \frac{1}{2\pi} \int_{-\pi}^{\pi} \sqrt{m^2+4\sin^{2}{\left( \frac{p}{2}\right) }} e^{ipn} dp \\ &=2  (2+m^2) \bra{0}\phi_0\phi_{n}\ket{0} - 2\bra{0}\phi_0\phi_{n+1}\ket{0} - 2\bra{0}\phi_0\phi_{n-1}\ket{0} \\ &= -\frac{1}{2\sqrt{\pi}(2+m^2)^{n-1/2}}\frac{\Gamma\left[ -\frac{1}{2} +n \right]}{\Gamma\left[ 1 +n \right]}  \pFq{2}{1}{\frac{-1+2n}{4}\,,\frac{1+2n}{4}}{1+n}{\frac{4}{(2+m^2)^2}}  \ \ \ .
\label{eq:seriessimp}    
\end{align}
Eq.~\eqref{eq:seriessimp} provides a compact representation utilizing a series expansion (Eq.~(15.2.2) in Ref.~\cite{NIST:DLMF}) of the $_{2}F_1$ hypergeometric function. From the positive definiteness of physical CMs~\cite{serafini2017quantum}, the block diagonal structure of $\sigma_{vac}$ leads to $K > 0$ and $K^{-1} > 0$. Principle submatrices of $K$ and $K^{-1}$, denoted as $K_{\mathbf{r}}$ and $\left(K^{-1}\right)_{\mathbf{r}}$ for mode list $\mathbf{r}$, satisfy the same relation, 
\begin{equation}
K_\mathbf{r} > 0 \quad , \quad \left(K^{-1}\right)_{\mathbf{r}} >0 \ \ \ .
\label{eq:referenceable}
\end{equation}
This follows from $K > 0$ and $K^{-1} > 0$ being equivalent to $v^{T} K v > 0$ and $v^{T} K^{-1} v > 0$ for all vectors $v$, including the specific choice of $v$ having support only in the subspace of principle submatrices. Because tracing modes in the Gaussian formalism is performed by isolating the CM principle submatrix for the retained modes~\cite{serafini2017quantum}, Eq.~\eqref{eq:referenceable} indicates that reduced field and conjugate-momentum CMs, e.g., relevant for a single or pair of detection region(s), $\sigma_{\phi,\mathbf{r}}$ and $\sigma_{\pi,\mathbf{r}}$, are also positive definite (PD).

\subsection{Semi-definiteness of region-region field- and momentum-space CMs}
\label{app:fieldintrooff}

In the thermodynamic limit of the latticized free scalar field vacuum with modes arranged as in Eq.~\eqref{eq:orderofcan}, region-region field-(momentum-)space CMs have positive-(negative-)semidefinite matrix properties. The following provides an analytic proof of these properties, Eq.~\eqref{eq:offblockmatcondition}, for the mixed state of disjoint vacuum regions.

With the ordering of canonical operators as in Eq.~\eqref{eq:orderofcan}, translation symmetry yields a mixed state CM of the form,
\begin{equation}
\sigma = \sigma_\phi \oplus \sigma_\pi = \begin{pmatrix}
\sigma_{\phi,A} & \sigma_{\phi,AB} \\
\sigma_{\phi,AB} & \sigma_{\phi,A} \\
\end{pmatrix} \oplus \begin{pmatrix}
\sigma_{\pi,A} & \sigma_{\pi,AB} \\
\sigma_{\pi,AB} & \sigma_{\pi,A} \\
\end{pmatrix} \ \ \ ,
\label{eq:vacsym}
\end{equation}
such that $\left(\sigma_{\phi,AB}\right)_{ij} = 2\bra{0}\phi_0\phi_{\Tilde{r}+i+j+1}\ket{0}$ and $(\sigma_{\pi,AB})_{ij} = 2\bra{0}\pi_0\pi_{\Tilde{r}+i+j+1}\ket{0}$, where $i$ and $j$ start from zero. Note that, the mode ordering established in Eq.~\eqref{eq:orderofcan} results in symmetric $\sigma_{\phi,AB}$ and $\sigma_{\pi,AB}$ in addition to the symmetric $\sigma_\pi$ and $\sigma_\phi$. 

Employing an integral representation (Eq.~(15.6.1) in Ref.~\cite{NIST:DLMF}) of the $_{2}F_1$ hypergeometric function in Eq.~\eqref{eq:fieldhyp},
\begin{equation}   
    \frac{1}{\Gamma\left[ 1 +n \right]} \pFq{2}{1}{\frac{1+2n}{4}\,,\frac{3+2n}{4}}{1+n}{\frac{4}{(2+m^2)^2}}= \frac{ \int_{0}^{1} t^{\frac{2n-1}{4}} (1-t)^{\frac{2n-3}{4}} \left(1-\frac{4}{(2+m^2)^2}t\right)^{-\frac{1+2n}{4}}dt}{\Gamma\left[ \frac{3+2n}{4} \right] \Gamma\left[ \frac{1+2n}{4} \right]} \ \ \ ,
    \label{eq:fieldfactor2}
\end{equation}
and the Legendre duplication formula, $\Gamma(z) \Gamma(z+1/2) = \Gamma(2z) \sqrt{\pi}/2^{2z-1}$, the region-region field-space CM may be written as an entrywise (Hadamard) product of two positive semidefinite (PSD) matrices, $\left(\sigma_{\phi,AB}\right)_{ij} \equiv q_{ij}  q'_{ij} $, where $q_{ij}$ and $q'_{ij}$ are defined by,
\begin{equation}
q_{ij}  \equiv \frac{2^{\tilde{r}+i+j+1/2}}{\pi(2+m^2)^{\tilde{r}+i+j+3/2}},\quad q'_{ij} \equiv \int_{0}^{1} t^{\frac{2\tilde{r}+2i+2j+1}{4}} (1-t)^{\frac{2\tilde{r}+2i+2j-1}{4}} \left(1-\frac{4}{(2+m^2)^2}t\right)^{-\frac{2\tilde{r}+2i+2j+3}{4}} dt \ \ \ .
\label{eq:fieldfactor}
\end{equation}
To see the PSD property of these matrices, note that $q$ is proportional to the outer product of $\left(1, \frac{2}{2+m^2},\left(\frac{2}{2+m^2}\right)^2,...\right)^{T}$, and that $q'_{ij}$ can be reformulated as a matrix of inner products of linearly independent functions proportional to, $t^{\frac{j}{2}} (1-t)^{\frac{j}{2}} \left(1-\frac{4}{(2+m^2)^2}t\right)^{-\frac{j}{2}}$. From the Schur product theorem~\cite{schurproductoriginal,hornmatrixana} (If Q,P are PSD Hermitian matrices, the entrywise product $Q \circ P$ is also PSD), the entrywise product of $q$ and $q'$ is PSD, hence $\sigma_{\phi,AB} \geq 0$, as stated in Eq.~\eqref{eq:offblockmatcondition}.

A similar line of reasoning that leads to $\sigma_{\pi,AB} \leq 0$ can be realized from Eq.~\eqref{eq:seriessimp}, where $\left(\sigma_{\pi,AB}\right)_{ij}  \equiv -p_{ij} p'_{ij} p''_{ij}$. While $p_{ij}$ and $p'_{ij}$ result from a decomposition similar to that of Eq.~\eqref{eq:fieldfactor}, an additional factor reads,
\begin{equation}
  p''_{ij} = \frac{1}{2\tilde{r}+2i+2j+1} = \int_{0}^{\infty} e^{-(2\tilde{r}+2i+2j+1)t} dt \ \ \ ,
\end{equation}
which can be reformulated as a matrix of inner products of linearly independent functions proportional to $e^{-2jt}$. Thus, $p''$ is PSD and a second application of the Schur product theorem yields $\sigma_{\pi,AB} \leq 0$, as stated in Eq.~\eqref{eq:offblockmatcondition}.

Simplifying matrix inversions in the symmetry context of Eq.~\eqref{eq:vacsym} with the block-diagonalizing similarity transform $S_{0} \equiv \frac{1}{\sqrt{2}}\begin{pmatrix} \mathbb{I}_{d} & \mathbb{I}_{d} \\ -\mathbb{I}_{d} & \mathbb{I}_{d} \end{pmatrix}$ yields the following expression for the region-region block of the momentum space CM,
\begin{equation}
\sigma_{\pi,AB} = \frac{\left(\left(\sigma_{\pi}^{-1}\right)_{A}+\left(\sigma_{\pi}^{-1}\right)_{AB}\right)^{-1}-\left(\left(\sigma_{\pi}^{-1}\right)_{A}-\left(\sigma_{\pi}^{-1}\right)_{AB}\right)^{-1}}{2} \leq 0 \quad \Longrightarrow \quad \left(\sigma_{\pi}^{-1}\right)_{AB} \geq 0 \ \ \ .
\label{eq:offblockmat}
\end{equation}
The final PSD property for the region-region block of the inverse momentum-space CM results from the extension of Eq.~\eqref{eq:referenceable} to $S_0 \sigma_\pi^{-1} S_0^T > 0$, allowing the inverse inequality to be inverted.

In systems such as disjoint regions of the lattice scalar field vacuum in higher dimensions~\cite{NKnegativitysphere} and the Gaussian approximation of local ion-chain axial motional modes symmetrically distributed in a quadratic trap~\cite{Retzker:2005dve,NKphononatural}, semi-definiteness of region-region field(position)- and momentum-space CMs i.e., Eq.~\eqref{eq:offblockmatcondition}, are also numerically observed. This indicates opportunity for extending the analytic entanglement properties presented in Appendix B to a broad class of physical quantum systems.

\section{Features of entanglement between disjoint vacuum regions}
\label{app:struc}

\subsection{Negativity as a necessary and sufficient entanglement measure}
\label{app:struc1}

This section proves that negativity is necessary and sufficient for identifying entanglement between two many-body disjoint regions of scalar field vacuum. The derivation utilizes properties of field- and momentum-space CMs, e.g., Eq.~\eqref{eq:offblockmatcondition} as derived in Appendix~\ref{app:fieldintro}.

Consider the PT CM of two regions in the free (massive or massless) scalar field,
\begin{equation}
    \Tilde{\sigma} = \Lambda \sigma \Lambda =\sigma_\phi \oplus \tilde{\sigma}_\pi = \begin{pmatrix}
\sigma_{\phi,A} & \sigma_{\phi,AB} \\
\sigma_{\phi,AB} & \sigma_{\phi,A} \\
\end{pmatrix} \oplus \begin{pmatrix}
\sigma_{\pi,A} & -\sigma_{\pi,AB} \\
-\sigma_{\pi,AB} & \sigma_{\pi,A} \\
\end{pmatrix} \ \ \ ,
\label{eq:PTCM}
\end{equation}
where $\tilde{\sigma}_\pi > 0$ as it can be obtained from $\sigma_\pi$ via similarity transformation. With vanishing $\phi\pi$-mixed matrix elements, the Peres-Horodecki criterion can thus be simplified via the Schur complement of $\tilde{\sigma}+i \Omega$ as the following necessary condition for a state to be separable,
\begin{equation}
\sigma_\phi - \left( \tilde{\sigma}_\pi\right)^{-1} \geq 0 \quad \Longleftrightarrow \quad \begin{cases}
      \sigma_{\phi,A} + \sigma_{\phi,AB} - \left(\sigma_{\pi}^{-1}\right)_{A} + \left(\sigma_{\pi}^{-1}\right)_{AB} \geq 0\\
      \sigma_{\phi,A} - \sigma_{\phi,AB} - \left(\sigma_{\pi}^{-1}\right)_{A} - \left(\sigma_{\pi}^{-1}\right)_{AB} \geq 0
    \end{cases} \ \ \ .
\label{eq:ncontribute}
\end{equation}
Reducing to the pair of half-sized PSD relations on the right utilizes symmetry~\footnote{Note that the inverse of a symmetric matrix with symmetric off-diagonal blocks is a matrix of the same form.} by similarity transforming the left hand side by $S_{0}$. The first inequality on the right of Eq.~\eqref{eq:ncontribute} is automatically satisfied given Eqs.~\eqref{eq:offblockmatcondition} and~\eqref{eq:offblockmat} along with the principle submatrix of the physicality condition $\sigma_\phi - \sigma_\pi^{-1} \geq 0$, which yields $\sigma_{\phi,A} - \left(\sigma_{\pi}^{-1}\right)_{A} \geq 0$. This corresponds to physicality of $\tilde{\sigma}_{\pm} \equiv (\sigma_{\phi,A} + \sigma_{\phi,AB}) \oplus (\sigma_{\pi,A}  -\sigma_{\pi,AB})$, which is always physical for the free scalar vacuum~\footnote{Note that any real PD matrix $M \equiv \begin{pmatrix}
M_{A} & M_{AB} \\
M_{AB} & M_{A}
\end{pmatrix}$ has a relation $(M_{A} \pm M_{AB})^{-1} = (M^{-1})_{A} \pm (M^{-1})_{AB}$.}. The second inequality is equivalent to the physicality condition of $\tilde{\sigma}_{\mp} \equiv (\sigma_{\phi,A} - \sigma_{\phi,AB}) \oplus (\sigma_{\pi,A}  + \sigma_{\pi,AB})$. A separable state with the same domain of physicality as $\tilde{\sigma}$ can thus be constructed from $\tilde{\sigma}_{\mp}$ as,
\begin{equation}
    \sigma^{sep} \equiv \begin{pmatrix}
\sigma_{\phi,A} - \sigma_{\phi,AB} & 0 \\
0 & \sigma_{\phi,A} - \sigma_{\phi,AB} \\
\end{pmatrix} \oplus \begin{pmatrix}
\sigma_{\pi,A} + \sigma_{\pi,AB} & 0 \\
0 & \sigma_{\pi,A} + \sigma_{\pi,AB} \\
\end{pmatrix}   \ \ \ ,
\label{eq:sepform}
\end{equation}
which respects $\sigma \geq \sigma^{sep}$, again utilizing Eq.~\eqref{eq:offblockmatcondition}. An alternate way to appreciate the separable solution of Eq.~\eqref{eq:sepform} that guarantees $\sigma \geq \sigma^{sep}$ is through a subtraction of classical noise with the form $Y = \sigma- \sigma^{sep} = \sigma_{\phi,AB} \begin{pmatrix}
\mathbb{I} & \mathbb{I} \\
\mathbb{I} & \mathbb{I}
\end{pmatrix}
 \oplus
\sigma_{\pi,AB} \begin{pmatrix}
-\mathbb{I} & \mathbb{I} \\
\mathbb{I} & -\mathbb{I}
\end{pmatrix}$ from the CM, which is the minimum allowable noise retaining symmetries and simple block diagonal structure~\footnote{A CM is separable if and only if there exists two CMs $\sigma_{A}$ and $\sigma_{B}$ such that $\sigma \geq \sigma_{A} \oplus \sigma_{B}$~\cite{Gaussianboundent}. If the CM is additionally AB symmetric and $\phi\pi$-uncorrelated, $\sigma \geq \sigma_{A} \oplus \sigma_{A}$ with no $\phi \pi$-mixing is always available~\cite{WolfGEOF}.}. Physically speaking, if the spacelike vacuum negativity vanishes, the many-body $\sigma$ can be created from $\sigma^{sep}$ through a Gaussian quantum channel. Providing analytic clarity to the numerical observation of Ref.~\cite{NKentsphere}, violating the Peres-Horodecki criterion (non-zero negativity) is thus promoted to both a necessary and sufficient condition for the presence of entanglement between symmetric disjoint regions of the free scalar field vacuum.

If PSD conditions as in Eq.~\eqref{eq:offblockmatcondition} are not satisfied, the Peres–Horodecki criterion is here further observed to be necessary and sufficient for \textit{asymmetric} disjoint regions of the free scalar field vacuum and local axial motional modes of trapped ion-chains~\cite{NKphononatural}, which can be numerically identified by the separability flow of Ref.~\cite{Giedkesepflow}. In addition, the necessary and sufficiency extends to classifying the minimum noise filtered (MNF) halo as separable or entangled in these systems~\footnote{Beyond the separability flow~\cite{Giedkesepflow}, an MNF halo that is separable is observed to satisfy serveral stronger versions of the Peres-Horodecki criterion that lead to separable state identification. For example, $\sigma_{\phi,A} - \sigma_{\phi,AB}^{+} - \sigma_{\phi,AB}^{-} - \left(\sigma_{\pi}^{-1}\right)_{A} - \left(\sigma_{\pi}^{-1}\right)_{AB}^{+} - \left(\sigma_{\pi}^{-1}\right)_{AB}^{-} \geq 0$ with $M \equiv M^{+} - M^{-}$, and $M^{\pm} \geq 0$ identified through the eigendecomposition of $M$, corresponding to positive and negative eigenvalues respectively. A separable state with the same domain of physicality can be similarly constructed as $\sigma^{sep} \equiv \begin{pmatrix}
\sigma_{\phi,A} - \sigma_{\phi,AB}^{+} - \sigma_{\phi,AB}^{-} & 0 \\
0 & \sigma_{\phi,A} - \sigma_{\phi,AB}^{+} - \sigma_{\phi,AB}^{-} \\
\end{pmatrix} \oplus \begin{pmatrix}
 \left( \left(\sigma_{\pi}^{-1}\right)_{A} + \left(\sigma_{\pi}^{-1}\right)_{AB}^{+} + \left(\sigma_{\pi}^{-1}\right)_{AB}^{-} \right)^{-1} & 0 \\
0 & \left( \left(\sigma_{\pi}^{-1}\right)_{A} + \left(\sigma_{\pi}^{-1}\right)_{AB}^{+} + \left(\sigma_{\pi}^{-1}\right)_{AB}^{-} \right)^{-1} \\
\end{pmatrix}$, which respects $\sigma \geq \sigma^{sep}$.}, a critical step for determining whether a gap is present between the negativity and its requirement for state formation~\cite{gao2024partialtransposeguided}. As such, opportunities exist to extend the necessary and sufficiency of the Peres-Horodecki criterion for identifying entanglement to a broader class of many-body quantum states.

This section has analytically identified two physical mixed many-body contexts for which a necessary and sufficient entanglement criterion is computationally available. For separable states in these contexts, no additional resources beyond the logarithmic negativity calculation are needed. For inseparable states, these capabilities applied to the MNF halo offers significant computational advantage in alleviating the high numerical precisions that can be required for generic Gaussian many-body separability determinations~\cite{Giedkesepflow} to classify operational entanglement properties~\cite{gao2024partialtransposeguided}.

\subsection{Sufficient condition for \texorpdfstring{$\mathcal{N}$}{N}-SOL entanglement classification}
\label{app:stru2}

The $\mathcal{N}$-SOL entanglement class~\cite{gao2024partialtransposeguided} contains the subset of many-body mixed Gaussian quantum states for which the entirety of the logarithmic negativity can be sequentially extracted through local operations in the simple form of noisy two-mode $(1_A \times 1_B)$ squeezed vacuum states. While membership in this class can be generically determined by assessing the symplectic orthonormality of local region components in the system's PT symplectic eigenvectors, the following shows that the CM structure in Eq.~\eqref{eq:offblockmatcondition} leads directly to a sufficient condition for $\mathcal{N}$-SOL entanglement classification in AB-symmetric states. This promotes the $\mathcal{N}$-SOL entanglement classification of disjoint regions in the continuum scalar field vacuum from a numerical observation~\cite{gao2024partialtransposeguided} to an analytic result.  

To illuminate the $\mathcal{N}$-SOL classification, consider the symplectic transformation, $\left(S_0 \oplus S_0\right)$, of the PT CM,
\begin{equation}
(S_{0}\oplus S_{0})   \Tilde{\sigma} (S_{0}\oplus S_{0})^{T} = \begin{pmatrix}
\sigma_{\phi,A} + \sigma_{\phi,AB} & 0 \\
0 & \sigma_{\phi,A} - \sigma_{\phi,AB} \\
\end{pmatrix} \oplus \begin{pmatrix}
\sigma_{\pi,A}  -\sigma_{\pi,AB} & 0 \\
0 & \sigma_{\pi,A} + \sigma_{\pi,AB} \\
\end{pmatrix}  \ \ \  ,
\label{eq:consolidation}
\end{equation}
which leaves the PT symplectic eigenvalues invariant. Note that, up to reordering of the canonical operators, $(S_0 \oplus S_0)\tilde{\sigma}(S_0 \oplus S_0)^T \stackrel{\text{reorder}}{=}  \tilde{\sigma}_{\pm} \oplus \tilde{\sigma}_{\mp}$. Because each diagonal block of Eq.~\eqref{eq:consolidation} is real and positive, there exists a normal mode decomposition~\cite{williamsonnormalform,serafini2017quantum} for $\tilde{\sigma}_{\pm}$ and $\tilde{\sigma}_{\mp}$. Because $\tilde{\sigma}_{\pm}$ is always physical, the negativity contributing
subspace ($\mathcal{V}_{\mathcal{N}}$) is isolated to the latter. As a result, the local $\mathcal{V}_{\mathcal{N}}$ subspace will naturally have symplectic orthogonality, and thus $\sigma$ is a member of the $\mathcal{N}$-SOL entanglement class~\cite{gao2024partialtransposeguided}. From this result, the proposal in footnote 3 of Ref.~\cite{gao2024partialtransposeguided} for calculating the entanglement consolidating transformation in this symmetric context can be understood as the symplectic transformation governed by the normal mode decomposition of $\tilde{\sigma}_{\mp} \stackrel{\text{reorder}}{=} \tilde{\sigma}_A - \tilde{\sigma}_{AB}$, diagonalizing $\mathcal{V}_\mathcal{N}$. It is worth mentioning that Eq.~\eqref{eq:offblockmatcondition} is a sufficient but not a necessary condition for $\mathcal{N}$-SOL entanglement classification. States satisfying variants of Eq.~\eqref{eq:offblockmatcondition}, e.g., $\sigma_{\phi,AB} \leq 0$ and $\sigma_{\pi,AB} \geq 0$ obtained by swapping the field- and momentum-space CMs, also exhibit the above entanglement features.

\subsection{Reduction of computational resources}
\label{app:stru3}

Because the $\mathcal{V_{\mathcal{N}}}$ of $\sigma$ is completely captured by the symplectic eigenvalues less than one of $\tilde{\sigma}_{\mp}$, this combination can be used to calculate the negativity and symplectic transformation that diagonalizes the PT CM. With Eq.~\eqref{eq:consolidation}, the $n_{-} \leq d$ PT symplectic eigenvalues less than one can be calculated from,
\begin{equation}
\sqrt{\text{spec}\big( ( \sigma_{\phi,A} - \sigma_{\phi,AB})(\sigma_{\pi,A} + \sigma_{\pi,AB})\big)} \ \ \ ,
\label{eq:b5}
\end{equation}
where $\text{spec}\left( \cdot \right)$ denotes eigenvalues of the corresponding matrix. This relation provides a second numerically advantageous halving, requiring only a $d$-dimensional calculation to determine the $\mathcal{V}_\mathcal{N}$ subspace of a $4d$-dimensional CM. 

The momentum and field space row vectors of the symplectic transformation for $\mathcal{V_{\mathcal{N}}}$ symplectic eigenvalues may be expressed as AB-antisymmetric row vectors $\langle v_\pi | \propto \begin{pmatrix} |v_\pi^{half}\rangle \\ -|v_\pi^{half}\rangle \end{pmatrix}^T$ and $\langle v_\phi | \propto \begin{pmatrix} |v_\phi^{half}\rangle \\ -|v_\phi^{half}\rangle \end{pmatrix}^T$ with,
\begin{equation}
    | v_\pi^{half}\rangle = \text{vec}\big( ( \sigma_{\phi,A} - \sigma_{\phi,AB})(\sigma_{\pi,A} + \sigma_{\pi,AB})\big), \quad | v_\phi^{half}\rangle =  (\sigma_{\pi,A} + \sigma_{\pi,AB}) | v_\pi^{half}\rangle \ \ \ ,
    \label{eq:reddim}
\end{equation}
where $\text{vec}\left(\cdot\right)$ denotes the corresponding eigenvectors. Following procedures in Appendix B2 of Ref.~\cite{gao2024partialtransposeguided}, an additional symplectic normalization and single-mode squeezing operation are applied to $\ket{v_{\pi}}$ and $\ket{v_{\phi}}$ as overall normalization factors such that $\bra{v_{\phi}}\ket{v_{\pi}}=1$ and $    \bra{v_{\phi}}\sigma_{\phi}\ket{v_{\phi}} = \bra{v_{\pi}} \tilde{\sigma}_\pi \ket{v_{\pi}} $. Stacking these row vectors as $\tilde{S}_{\mathcal{V}_{\mathcal{N}}} = \mathbf{v}_\phi \oplus \mathbf{v}_\pi$ produces the $(2d \times 4d)$-dimensional $\mathcal{V}_{\mathcal{N}}$ subspace of the normal form symplectic, $\tilde{S}_{\mathcal{V}_\mathcal{N}} \tilde{\sigma} \tilde{S}_{\mathcal{V}_\mathcal{N}}^T = \Tilde{D}_{\mathcal{V}_\mathcal{N}}$ with $\tilde{S}_{\mathcal{V}_\mathcal{N}} \Omega_{2d} \tilde{S}_{\mathcal{V}_{\mathcal{N}}}^T  = \Omega_{d}$.  
By similarly constructing $\tilde{S}_{\mathcal{V}_{\Nslash}}$ through $\tilde{\sigma}_{\pm}$ from AB-symmetric row vectors $\langle v_\pi| \propto \left(\langle v_\pi^{half}|  \langle v_\pi^{half}|\right)$ and $\langle v_\phi| \propto \left(\langle v_\phi^{half}|  \langle v_\phi^{half}|\right)$ with $| v_\pi^{half}\rangle = \text{vec}\big(\left( \sigma_{\phi,A}+\sigma_{\phi,AB}\right)\left( \sigma_{\pi,A} - \sigma_{\pi,AB} \right)\big) $ and $| v_\phi^{half}\rangle =  \left( \sigma_{\pi,A} - \sigma_{\pi,AB} \right) | v_\pi^{half}\rangle$, the entire normal form symplectic may be produced through these reduced-dimensionality techniques.

The numerical simplification discussed in this section also applies to the creation of the local symplectic transformations capable of consolidating the negativity of  many-body $\mathcal{N}$-SOL states, e.g., the pairs of scalar field vacuum regions, into a tensor product of two-mode $(1_A \times 1_B)$ pairs between the regions, as discussed in Refs.~\cite{NKcorehalo,gao2024partialtransposeguided}. As shown in Ref.~\cite{gao2024partialtransposeguided} and dimensionally reduced in Eq.~\eqref{eq:reddim}, this local transformation can be built from the A- and B-space of the $\mathcal{V}_\mathcal{N}$ row vectors diagonalizing the PT CM, $\tilde{S}_{\mathcal{V}_{\mathcal{N}}}$. After implementing such transformations, additional single-mode symplectic transformations may be applied, as in Fig.~\ref{fig:localsprowvec}, which do not alter the consolidated entanglement.

\section{Maximum \texorpdfstring{$(1_A \times 1_B)$}{(1A x 1B)}-mode extractable entanglement}
\label{app:opt}

As seen in Eq.~\eqref{eq:negg}, the largest contribution to the logarithmic negativity is governed by the smallest PT symplectic eigenvalue, $\Tilde{\nu}_{min}$. For the entanglement between regions of the massive or massless free scalar field vacuum, the hierarchy among these contributions is exponential. As an entanglement monotone~\cite{computablemeasure,PlenioLogarithmic}, the logarithmic negativity is known to be non-increasing under local operations, including the tracing of modes in a many-body system. This leads the amount of entanglement that can be locally consolidated into a $(1_A \times 1_B)$ Hilbert subspace to be upper bounded by the logarithmic negativity of the multi-mode state. However, a stronger upper bound is here shown to be governed by the smallest PT symplectic eigenvalue of the multi-mode state.  As such, states with $n_- > 1$ negativity-contributing PT symplectic eigenvalues have a fundamental incompressibility in their quantum correlations, i.e., have a portion of their entanglement that eludes detection by a single pair of single-mode detectors. This upper bound of $(1_A \times 1_B)$ entanglement extraction is saturated by the consolidation techniques~\cite{NKcorehalo,gao2024partialtransposeguided} that lead to the presented collective operator detection profiles.

\subsection{Extremization representation of \texorpdfstring{$\Tilde{\nu}_{min}$}{vmin}}
\label{app:c1}

The following establishes the minimum expectation value of $\tilde{\sigma}$ with respect to symplectically orthonormal vectors as a proxy for the minimum PT symplectic eigenvalue. Beyond the present application, these procedures are valid for any real PD matrix.

As a Hermitian operator with an associated spectral decomposition, expectation values of $\tilde{\sigma}$ with respect to orthonormal vectors are well known to be bounded by its spectrum~\cite{hornmatrixana}. Such relationships do not extend, however, to the symplectic spectrum and expectation values with respect to symplectically orthonormal vectors, i.e., those pairwise satisfying,
\begin{equation}
\langle v^{(1)} | \tilde{\sigma} | v^{(1)} \rangle =  \langle v^{(2)} | \tilde{\sigma} | v^{(2)} \rangle  \quad , \quad \langle v^{(1)} | \Omega | v^{(1)} \rangle = \langle v^{(2)} | \Omega | v^{(2)} \rangle = 0 \quad , \quad \langle v^{(1)} | \Omega | v^{(2)} \rangle = 1 \ \ \ .
\label{eq:opt1}
\end{equation}
The $\tilde{S}$ symplectic operator that diagonalizes the PT CM to normal form, $\tilde{S} \tilde{\sigma} \tilde{S}^T = \Tilde{D}$ where $\Tilde{D}$ is a doubly degenerate diagonal matrix of PT symplectic eigenvalues, provides an initial symplectic basis. Denoting the two symplectically orthonormal $\tilde{S}$ row vectors spanning each of these degenerate subspaces as $|\tilde{\nu}_{j}^{(1)}\rangle$ and $|\tilde{\nu}_{j}^{(2)}\rangle$, this degeneracy leads to $\langle \tilde{\nu}_j^{(1)} | \tilde{\sigma} |\tilde{\nu}_j^{(1)}\rangle = \langle \tilde{\nu}_j^{(2)} | \tilde{\sigma} |\tilde{\nu}_j^{(2)}\rangle = \tilde{\nu}_j$, where $\tilde{\nu}_j$ is the $j$-th PT symplectic eigenvalue, in ascending order.
When the CM has no $\phi\pi$-mixed matrix elements, it is common to construct these pairs to be composed of vectors isolated to the position and conjugate-momentum space, respectively.

Built from the normal-form symplectic basis, consider, for example, the following family of symplectic vector pairs deviating from those associated with the lowest PT symplectic eigenvalue,
\begin{equation}
|w^{(1)}\rangle = |\tilde{\nu}_{1}^{(1)} \rangle + \alpha |\tilde{\nu}_i^{(1)}\rangle \quad , \quad 
|w^{(2)}\rangle = |\tilde{\nu}_{1}^{(2)} \rangle + \alpha \sqrt{ \frac{\tilde{\nu}_i}{\tilde{\nu}_j} } |\tilde{\nu}_j^{(2)}\rangle  \ \ \  ,
\end{equation} 
where $\alpha$ is a free real parameter. With $i,j \in \{2, ..., 2d\}$ and $ i \neq j$, this form continues to satisfy the Eq.~\eqref{eq:opt1} symplectic orthonormality conditions. This family is observed to saturate the global bounds~\footnote{The same conclusion of Eq.~\eqref{eq:opt3} can be obtained from the general approach of Lagrange multipliers, where stationary values satisfy the eigenvalue equation of $-\Omega\Tilde{\sigma}\Omega\Tilde{\sigma}$. Since the expectation value of the PD operator $\tilde{\sigma}$ is bounded from below, the global minimum is the smallest in the eigenvalue spectrum of $\sqrt{-\Omega\Tilde{\sigma}\Omega\Tilde{\sigma}}$. Subsequent connection to $\tilde{S}$ row vectors is discussed, for example, in Appendix B2 of Ref.~\cite{gao2024partialtransposeguided}.},
\begin{equation}
\min_{|v^{(1)}\rangle, |v^{(2)}\rangle} \langle v^{(1)} | \tilde{\sigma} |v^{(1)}\rangle = \tilde{\nu}_1  
\qquad , \qquad 
\max_{|v^{(1)}\rangle, |v^{(2)}\rangle} \langle v^{(1)} | \tilde{\sigma} |v^{(1)}\rangle = \infty
\label{eq:opt3}   \ \ \ .
\end{equation}
In particular, the expectation value $\langle w^{(1)} | \tilde{\sigma} |w^{(1)}\rangle = \tilde{\nu}_1+\alpha^2 \tilde{\nu}_i \geq \tilde{\nu}_1$ is minimized by $\alpha = 0$ as $\tilde{\nu}_j > 0\ \forall j$. Thus, the minimization of the $\tilde{\sigma}$ expectation value subject to Eq.~\eqref{eq:opt1} is the smallest PT symplectic eigenvalue itself. With a freedom present in the vector normalization while maintaining the symplectic normalization, the expectation value is shown also to be unbounded from above, as $\alpha$ can be made arbitrarily large.

\subsection{Monotonicity of \texorpdfstring{$\Tilde{\nu}_{min}$}{vmin} under collective operator identification}
\label{app:profiles}

In the optimization language of the previous section, the minimum possible PT symplectic eigenvalue of a $(1_A \times 1_B)$ pair of locally collective modes is here shown to be equivalent to that of the many-body system from which the collective modes have been selected.  As such, it is the minimum PT symplectic eigenvalue, not the total negativity, that governs the maximum amount of entanglement that could be transferred into a pair of single-mode detectors. 

Collective operators are defined by real linear combinations of canonical operators in each local region, i.e., the phase-space operators for a collective mode may be expressed as in Eq.~\eqref{eq:generalizeZpapaerprofiles} of the main text. Stacking the profile components in the ordering of Eq.~\eqref{eq:orderofcan} as,
\begin{equation}
|f_{X,A}\rangle = \begin{pmatrix}
|f^X_{\phi,A}\rangle \\ \vec{0}_d \\ | f^X_{\pi,A}\rangle \\ \vec{0}_d
\end{pmatrix} \quad , \quad  |f_{X,B} \rangle = \begin{pmatrix}
\vec{0}_d \\ |f^X_{\phi,B}\rangle \\ \vec{0}_d  \\ | f^X_{\pi,B}\rangle 
\end{pmatrix}  \ \ \ ,
\label{eq:ABembedding}
\end{equation}
with $X \in \{\Phi, \Pi\}$, the normalization and $\mathfrak{Im}\left(f\right) = 0$ translates naturally to symplectic orthonormality conditions $\langle f_{\Phi,A} | \Omega | f_{\Pi,A}\rangle = 1$ and $\langle f_{\Phi,A} | \Omega | f_{\Phi,A} \rangle = \langle f_{\Pi,A} | \Omega | f_{\Pi,A} \rangle = 0$ for both the A- and B-space. Beyond the constructions of Refs.~\cite{onebeforeZpaper,Zpaper,missep1PRD2023,agullo2024multimodenaturespacetimeentanglement}~\footnote{Constraining Eq.~\eqref{eq:generalizeZpapaerprofiles} to equal $\phi\pi$-uncorrelated profiles in the field and conjugate-momentum space, i.e., $f^\Phi_\phi = f^\Pi_\pi$ and $f^\Phi_\pi = f^\Pi_\phi = 0$, recovers the subset of collective operators explored in Refs.~\cite{onebeforeZpaper,Zpaper,missep1PRD2023,agullo2024multimodenaturespacetimeentanglement}.}, the general form of Eq.~\eqref{eq:generalizeZpapaerprofiles} allows local $\phi\pi$-mixing and unequal profiles in the field and conjugate-momentum space. In the present application, the latter extension is found to be crucial for detecting spacelike bipartite entanglement at long distances in the free scalar field vacuum.

With CM matrix elements calculated as symmetric second-order operator expectation values, $\sigma = \text{Tr} \left[ \rho \left\{ \hat{\boldsymbol{r}}, \hat{\boldsymbol{r}}^{T}\right\}\right]$ with $\bar{\mathbf{r}} = 0$, linearity and the action of the trace in the operator space leads to,
\begin{equation}
    S\sigma S^{T}= \mathrm{Tr} \left[ \rho \left\{ S\hat{\boldsymbol{r}},\hat{\boldsymbol{r}}^{T} S^{T} \right\}   \right] \ \ \ .
    \label{eq:connectioncollective}
\end{equation}
As such, symplectic transformations $S$ of a CM are equivalent to performing symplectically orthonormal linear combinations of phase-space operators governed by $S$ row vectors.
Because the covariance matrix of a reduced density matrix is simply the submatrix associated with the remaining modes, the two-mode $(1_A \times 1_B)$ CM of a collective mode in the A-space with a collective mode in the B-space may be written as, 
\begin{equation}
\sigma_f = S_f \sigma S_f^T \ \ , \quad  S_f^T \equiv \begin{pmatrix}
|f_{\Phi,A} \rangle &  |f_{\Phi,B}\rangle & |f_{\Pi,A}\rangle & |f_{\Pi,B}\rangle
\end{pmatrix} \ \ \ ,
\end{equation}
with the $(4d \times 4)$-dimensional $S_f^T$ constructed as a series of column vectors governed by collective operator weights.

From Eq.~\eqref{eq:negg}, maximizing the two-mode logarithmic negativity entanglement measure, $\mathcal{N}_{(1_A \times 1_B)}$, is governed entirely by a minimization of the smallest PT symplectic eigenvalue. For a pair of local collective modes extracted from an entangled ($\mathcal{N}>0$) many-body state, this eigenvalue is minimized by,
\begin{equation}
\min_f 2^{-\mathcal{N}_{\left(1_A\times 1_B\right)}} = \min_{f, | v_f^{(1)}\rangle,| v_f^{(2)}\rangle}  \langle v_f^{(1)}|\Tilde{\sigma}_{f}| v_f^{(1)}\rangle  = \min_{|v^{(1)}\rangle,|v^{(2)}\rangle}  \langle v^{(1)}| \Tilde{\sigma}|v^{(1)}\rangle = \tilde{\nu}_1  \ \ \ ,
\label{eq:opt2}
\end{equation}
where $\tilde{\nu}_1$ is the minimum PT symplectic eigenvalue of the larger many-body quantum system. The first and third equalities are direct consequences of Appendix~\ref{app:c1}, relating minimum PT symplectic eigenvalues to minimum expectation values of the associated PT CM with respect to symplectically orthonormal vectors. The central equality arises from the collective operator CM relation $\tilde{\sigma}_f = \Lambda_1 S_f \Lambda_{d} \tilde{\sigma} \Lambda_{d} S_f^T \Lambda_1$. Beyond $\Lambda_d S_f^T \Lambda_1 |v_f^{(1)}\rangle$ satisfying the same double degeneracy and symplecticity as $|v^{(1)}\rangle$, this correspondence of domain can be seen as an arbitrary collective operator $|v_f\rangle$ expanded into the many-body local Hilbert spaces via an independently arbitrary local unitary (c.f. Eq.~(12) of Ref.~\cite{gao2024partialtransposeguided}) providing the same flexibility as the symplectic optimization. Optimizing over profile vectors in this context is therefore equivalent to symplectically orthonormal optimization in the many-body state. Connecting the first and last expressions of Eq.~\eqref{eq:opt2} therefore demonstrates that the minimum PT symplectic eigenvalue in $\mathcal{V}_\mathcal{N}$ is monotonically increasing under collective operator identification. When the larger quantum system is characterized by $n_- > 1$, this translates to a stronger upper bound (less than $\mathcal{N}$) on the amount of the many-body entanglement that can be compressed into two-mode.

As an explicit statement of the above result, Eq.~\eqref{eq:opt4} provides the profiles that saturate this stronger upper bound, and thus maximize the logarithmic negativity in the two-mode subspace of local collective operators,
\begin{equation}
S_f^T = \Lambda_d\begin{pmatrix}
\frac{|\tilde{\nu}^{(1)}_{1,A}\rangle }{\sqrt{\langle \tilde{\nu}^{(1)}_{1,A} | \Omega | \tilde{\nu}^{(2)}_{1,A}\rangle}} 
& 
\frac{|\tilde{\nu}^{(1)}_{1,B}\rangle }{\sqrt{\langle \tilde{\nu}^{(1)}_{1,B} | \Omega | \tilde{\nu}^{(2)}_{1,B}\rangle}} 
&
\frac{|\tilde{\nu}^{(2)}_{1,A}\rangle }{\sqrt{\langle \tilde{\nu}^{(1)}_{1,A} | \Omega | \tilde{\nu}^{(2)}_{1,A}\rangle}} 
&
\frac{|\tilde{\nu}^{(2)}_{1,B}\rangle }{\sqrt{\langle \tilde{\nu}^{(1)}_{1,B} | \Omega | \tilde{\nu}^{(2)}_{1,B}\rangle}} 
\end{pmatrix}\Lambda_1 \ \ \ ,
\label{eq:opt4}
\end{equation}
where the notation $|\tilde{\nu}^{(1)}_{1,X}\rangle$ indicates the $X\in \{A,B\}$ subspace of $|\tilde{\nu}_1^{(1)}\rangle$ is extracted and embedded into a $4d$ vector as performed in Eq.~\eqref{eq:ABembedding}. The normalization constants ensure symplectic normality of the profile vectors. Note that though $\mathcal{V}_\mathcal{N}$ row vectors exhibit AB-antisymmetry in symmetric field vacuum regions, both the AB-symmetric and AB-antisymmetric profiles are capable of extracting the dominant negativity contribution, i.e., the non-zero elements of the profile vectors can satisfy $|f_{X,A}\rangle = \pm|f_{X,B}\rangle$ for $X\in \{\Phi,\Pi\}$, where the two choices are related by a single-mode phase operation with angle $\pi$ applied to one of the collective modes. In general, there exists freedom by a full symplectic operation on each collective mode that generates a family of optimal profiles for $S_f$, e.g., allowing $\phi\pi$-mixing in addition to the interactions of Fig.~\ref{fig:localsprowvec} for extracting scalar field entanglement.

\section{From lattice to continuum entanglement}
\label{app:artfact}

\subsection{Properties of continuum entanglement}
\label{app:contlim}

Reviewing the continuous limit of the free scalar field vacuum, the Hamiltonian in terms of a one-dimensional spatial coordinate, $x$, may be written as,
\begin{equation}
    \hat{H} = \frac{1}{2} \int dx \left( \hat{\pi}(x)^{2} + \hat{m}^{2}\hat{\phi}(x)^{2} + (\nabla \hat{\phi}(x))^{2} \right) \ \ \ ,
    \label{eq:continuumH}
\end{equation}
where $\hat{\phi}$ and $\hat{\pi}$ are field and momentum operators satisfying equal-time CCRs, $\left[\hat{\phi}(x) , \hat{\pi}(x')\right] = i \delta(x-x')$. Upon lattice regularization with lattice spacing $a$, the $j$-th discretized field and momentum operators are defined as $\hat{\phi}_{j} \equiv \hat{\phi}(ja)$ and $\hat{\pi}_{j} \equiv \hat{\pi}(ja)$. Expressing lengths in units of this lattice spacing, a dimensionless Hamiltonian $H \equiv a \hat{H}$ can be formed with $m \equiv a\hat{m}$, $\phi_{j} \equiv \hat{\phi}_{j}$ and $\pi_{j} \equiv a\hat{\pi}_{j}$, recovering Eq.~\eqref{eq:descreteH}. The geometry of spacelike vacuum region pairs is subsequently captured by $d$ and $\tilde{r}$, the number of lattice sites in each region and between the two regions.

In a massive theory, the correlation length is set by the Compton wavelength scale, i.e., $\xi = 1/(a\hat{m}) = 1/m$. The physical configuration space depends only on the number of correlation lengths in each region, $m d$, and between the two regions, $m \tilde{r}$. In the present work, the continuous limit of an $(md, m\tilde{r})$ physical configuration is approached by fixing $a \equiv 1$, and increasing the number of lattice sites pixelating the geometry, $d$ and $\tilde{r}$, while correspondingly decreasing the individual oscillator mass, $m$~\footnote{Note that the present approach yields consistent results in the continuum with the approach of Refs.~\cite{scalarvacuumoriginal1,scalar1dextra,Zpaper}, where instead the lattice spacing is taken to zero and coupling constant is taken to one with fixed propagation velocity.}. In the scale-invariant massless theory, where the correlation length extends to infinity, the continuum limit is approached simply by increasing the number of oscillators pixelating physical configurations governed by the ratio $\tilde{r}/d$ as discussed, for example, in Refs.~\cite{scalarvacuumoriginal1,scalar1dextra,NKnegativitysphere,NKentsphere}. Because the derivations in Appendix~\ref{app:struc} are independent of the parameters $d$, $\tilde{r}$, and $m$, the presented entanglement features of vacuum disjoint regions persist into the continuum. In particular, the collective operator profiles converge to provide a pair of spatially smeared continuum modes through which local detectors can extract an exponentially dominant portion of the continuum spacelike entanglement.

\subsection{Entanglement decay and UV-IR connection}
\label{app:entsphere}

\begin{figure}[t!]

\centering
\includegraphics[width=.33\textwidth]{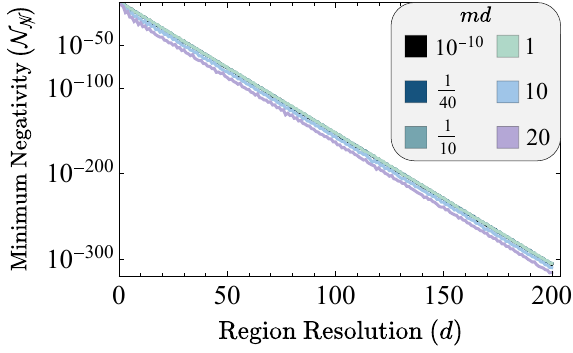}\hfill
\includegraphics[width=.33\textwidth]{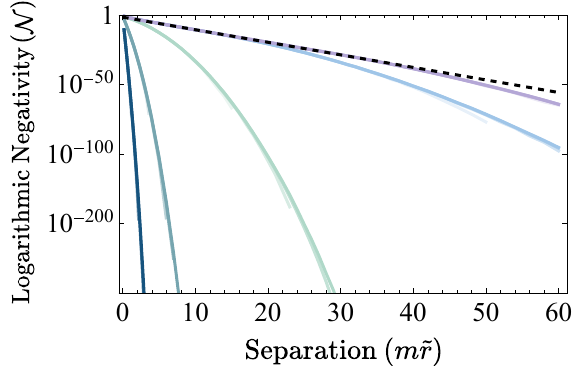}\hfill
\includegraphics[width=.33\textwidth]{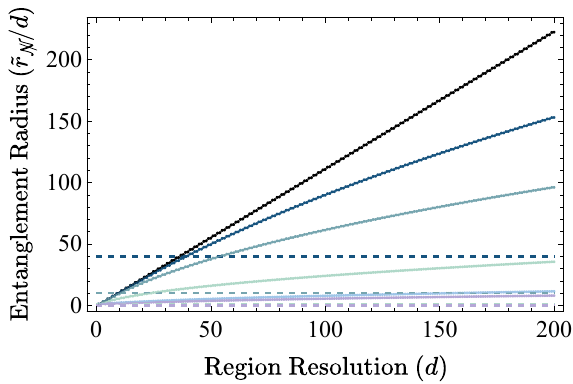}

\caption{For different choices of $md$, (left) the smallest non-zero value of the logarithmic negativity supported by the lattice regions, $\mathcal{N}_{\Nslash}$, as a function of region resolution, $d$. When $d \gtrsim 10$, the lines follow an exponential linear curve $\mathcal{N}_{\Nslash}  \sim e^{-\Lambda' d}$ with the same slope. (middle) Logarithmic negativity, $\mathcal{N}$, as a function of $m\tilde{r}$. To demonstrate convergence in the continuum, several digitizations are shown $d=\left( 50,100,150,200 \right)$, where lighter shade corresponds to smaller $d$. The dashed line to which the negativity is found to converge at increasing $md$ is informed by the $m\tilde{r}/md < 1$ regime of $md =20$, with a slope $\sim 2.1$. (right) Entanglement sphere radius, $(m \Tilde{r}_{\Nslash})/(m d)$, as a function of $d$. The calculation for $md=10^{-10}$ agrees with the massless result in Ref.~\cite{NKnegativitysphere}. Horizontal lines of corresponding color denote separations on the scale of one Compton wavelength ($m\tilde{r} = 1$ consistent with Ref.~\cite{Zpaper}). The curves fit square root functions of the form $a + b\sqrt{d+c}$. Consistent with the results of prior panels, this indicates that the entanglement sphere grows systematically with the UV pixelation beyond any finite separation, including the Compton wavelength scale.}
\label{fig:plots}

\end{figure}

By analyzing the decay and truncations of entanglement between regions of the massive lattice scalar field vacuum, this section determines that spacelike entanglement is available at separations that extend to infinity in the field continuum limit. Beyond numerical observation, this result is strengthened by a UV-IR connection between the long-distance entanglement and the short-distance properties of the field.

For fixed detector sizes in units of the Compton wavelength scale, $md$, the middle panel of Fig.~\ref{fig:plots} presents calculations of the spacelike logarithmic negativity (evaluated via Eq.~\eqref{eq:b5}) as a function of $m\tilde{r}$ dimensionless separation. The results agree with Ref.~\cite{scalar1dextra}, where $\mathcal{N}$ follows a quadratic exponential curve $\mathcal{N}\sim e^{-\left(m\tilde{r}\right)^2}$ at long distances and a linear exponential curve $\mathcal{N}\sim e^{-m\tilde{r}}$ indicated by the dashed line at shorter distances. The latter can also be seen in Fig.~\ref{fig:heatmap} where the exponentially decaying logarithmic negativity contours from the massless regime bend from radial to parallel at increasing $md$. Because the $md$-fixed ratio $m\tilde{r}/md$ at which the linear-to-quadratic transition occurs is found to increase with increasing mass, this provides evidence that the linear regime extends beyond the previously established~\cite{scalar1dextra} $m\tilde{r}/md = 1$ in the heavy mass limit. Finally, in the small mass limit, $\mathcal{N}$ is found to follow an exponential curve quadratic in $m\tilde{r}$ with a linear decay parameter scaling as $1/md$ and a quadratic contribution suppressed by mass, recovering the linear exponential decay as a function of $\tilde{r}/d$ in the massless limit~\cite{scalar1dextra}.

When a field is approximated with limited bandwidth, the spacelike entanglement supported by the approximation will be truncated at a finite distance. This can be seen, for example, in the lower-resolution calculations prior to continuum convergence in the middle panel of Fig.~\ref{fig:plots}. To express the size of the quantum mechanically coherent volume, the entanglement sphere of a latticized field is defined as the smallest dimensionless separation $\tilde{r}_{\Nslash}/d$ at (and beyond) which entanglement between disjoint regions vanishes. Any observable outside the entanglement sphere can be completely described by classical probability distributions.

According to Appendix~\ref{app:struc1}, the entanglement sphere of the scalar field vacuum coincides with the vanishing of logarithmic negativity. The smallest spacelike entanglement supported by the lattice, as shown in the left panel of Fig.~\ref{fig:plots}, is found to decay exponentially in $d$ with a decay parameter that is independent of mass. As such, the relationship between the UV truncation and smallest spacelike entanglement, $\mathcal{N}_{\Nslash} \sim e^{-\Lambda' d}$, is observed to persist from the massless regime. Beyond $md >1$, it is additionally observed that the entanglement decays also exponentially with mass, such that higher masses with more localized wavefunctions require higher momentum support for entanglement. This relationship can also be seen from the right panel where the entanglement sphere radius decreases at fixed region resolution, suggesting that long-distance quantum correlations probe shorter distance scales for fields of higher mass. These observations provide generalization to the UV-IR connection in Ref.~\cite{NKentsphere}~\footnote{Furthering this UV-IR connection, several modifications of the lattice action uncover an additional lattice artifact that distorts the $\mathcal{N}$-SOL entanglement classification~\cite{gao2024partialtransposeguided} shown in Appendix~\ref{app:stru2} to be present in the continuum.}.

In the right panel of Fig.~\ref{fig:plots}, the separation (in units of the detector size) between disjoint regions at which the negativity vanishes, $m\tilde{r}_{\Nslash}/md$, is presented as a function of region resolution in lattice units, $d$. With the functional forms of the second and third panels combining to produce that of the first, the quadratic exponential decay of logarithmic negativity with separation at large $m\tilde{r}/md$ corresponds to a squareroot expansion of the entanglement sphere at increasing resolution. Therefore, the continuum limit ($d \rightarrow \infty$) for a fixed physical detector size, $md$, yields $m \tilde{r}_{\Nslash}/md \rightarrow \infty$, i.e., systematically restoring accessible entanglement between disjoint regions at arbitrary distances in the massive field vacuum.

\end{document}